\documentstyle[epsfig,rotate]{aipproc2}
\begin{document}
\title{Relativistic mean-field description\\ of nuclei at the 
drip-lines}

\author{D. Vretenar$^{1,2}$, P. Ring$^{2}$, and G.A. Lalazissis$^{2}$}
\address{
$^{1}$ Physics Department, Faculty of Science, University of
Zagreb, Croatia\\
$^{2}$ Physik-Department der Technischen Universit\"at M\"unchen,
D-85748 Garching, Germany\\}

\maketitle

\begin{abstract}
We present a review of recent applications of the relativistic  
mean-field theory to the structure of nuclei close to the 
drip-lines. For systems with extreme isospin values, 
the relativistic Hartree-Bogoliubov model provides 
a unified and self-consistent description of mean-field and 
pairing correlations. The model has been applied in studies 
of structure phenomena that include: formation of neutron skin
and of neutron halos in light nuclei in the mass region above the s-d 
neutron shell, the strong isospin dependence of the effective
spin-orbit interaction and the resulting 
modification of surface properties, the suppression of the spherical
$N=28$ shell gap for neutron-rich nuclei and the related phenomenon
of deformation and shape coexistence, the proton drip-line in the 
spherical nuclei $14\leq Z \leq 28$, and ground-state proton 
radioactivity in the region of deformed nuclei $59 \leq Z \leq 69$.
\end{abstract}

\section*{Introduction}

Experimental and theoretical studies of exotic nuclei with
extreme isospin values present one of the most
active areas of research in nuclear physics. Experiments with
radioactive nuclear beams provide the opportunity to study
very short-lived nuclei with very large neutron to proton
ratios $N/Z$. New theoretical models and techniques are being
developed in order to describe unique phenomena in nuclei
very different from those usually encountered along the
line of stability. On the neutron-rich side in particular,
exotic phenomena include the weak binding of the outermost
neutrons, pronounced effects of the coupling between bound
states and the particle continuum, regions of nuclei
with very diffuse neutron densities, formation of the neutron
skin and halo structures. The modification of the
effective nuclear potential produces a
suppression of shell effects, the disappearance
of spherical magic numbers, and the onset of deformation
and shape coexistence. Isovector quadrupole
deformations are expected at the neutron drip-lines, and possible
low-energy collective isovector modes have been predicted.
Because of their relevance to the r-process in nucleosynthesis,
nuclei close to the neutron drip-line are also very important
in nuclear astrophysics. Their properties determine the 
astrophysical conditions for the formation of neutron-rich
stable isotopes. 

Proton-rich nuclei also display many interesting structure phenomena
which are important both for nuclear physics and astrophysics.
These nuclei are characterized by exotic ground-state decay
modes such as direct emission of charged particles and $\beta$-decays
with large Q-values. The properties of many proton-rich nuclei
should also play an important role in
the process of nucleosynthesis by rapid-proton
capture. In addition to decay properties (particle
emission, $\beta$-decay), of fundamental importance are
studies of atomic masses and separation energies, and especially the
precise location of proton drip-lines.

Particularly important are nuclear systems with an equal
number of proton and neutrons. Due to the high symmetry
between the proton and neutron degrees of freedom, it is
possible to study the details of the effective off-shell
$n-p$ interaction. Protons and neutrons occupy the same shell-model
orbitals and therefore these nuclei present unique systems in
which $(S=0, T=1)$ and $(S=1, T=0)$ pairing can be studied.
While for $A \leq 40$ the $N=Z$
nuclei are $\beta$-stable, in heavier systems the Coulomb
interaction drives the beta-stability line towards
neutron-rich isotopes. Heavier $N=Z$ nuclei approach the
proton drip-line and display a variety of structure phenomena:
shape coexistence, superdeformation, alignment of proton-neutron
pairs, proton radioactivity from highly excited states.

Some of the most
interesting examples of nuclear systems with large isospin values
are provided by the recent experimental results on the synthesis
and stability of the heaviest elements. The periodic system
has been extended with three new elements $Z=110, 111, 112$.
These elements are found beyond the macroscopic limit of nuclear 
stability and are stabilized only by quantal shell effects. 
All the heaviest elements found recently are believed
to be well deformed. Still heavier and more neutron-rich elements are
expected to be spherical and stabilized by shell effects,
which strongly influence the spontaneous fission and
alpha-decay half-lives.

While an impressive amount of experimental data on properties of exotic
nuclei has been published in the last decade, the development of theoretical
models has proceeded at a somewhat slower pace. Significant progress
has been reported in the development of mean-field theories and
theoretical models which use effective interactions to describe low
energy nuclear states. Calculations
based on modern Monte Carlo shell-model techniques provide a
relatively accurate
description of the structure of light and medium-mass exotic nuclei.
For medium-heavy and heavy systems the only viable approach at present are
large scale self-consistent mean-field calculations
(Hartree-Fock, Hartree-Fock-Bogoliubov, relativistic mean-field). However, 
the present accuracy of model predictions, in general, is not comparable
with the precision of modern experimental research facilities,
and there are many serious technical difficulties in the calculation
of properties of complex nuclear systems at the drip-lines.

Models based on quantum hadrodynamics \cite{SW.97} provide a framework 
in which the nuclear system is described by interacting baryons and mesons.
In comparison with conventional non relativistic descriptions,
relativistic models explicitly include mesonic
degrees of freedom and consider the nucleons as Dirac particles.
A variety of nuclear phenomena have been described in the
relativistic framework: nuclear matter, properties of finite
spherical and deformed nuclei, hypernuclei, neutron stars,
nucleon-nucleus and electron-nucleus
scattering, relativistic heavy-ion collisions.
In particular, relativistic models
based on the mean-field approximation have
been successfully applied in the description of properties
of spherical and deformed $\beta$-stable nuclei,
and more recently in studies of exotic nuclei far from the
valley of beta stability.

\section*{The relativistic Hartree-Bogoliubov model}

Detailed properties of nuclear matter and of finite nuclei
along the $\beta$-stability line have been very
successfully described with relativistic
mean-field models \cite{Rin.96}. The nucleus is described as 
a system of Dirac nucleons, which interact in a relativistic covariant 
manner through the exchange of virtual mesons:~the isoscalar
scalar $\sigma$-meson, the isoscalar vector $\omega$-meson
and the isovector vector $\rho$-meson. The effective
Lagrangian density of quantum hadrodynamics reads
\begin{eqnarray}
{\cal L}&=&\bar\psi\left(i\gamma\cdot\partial-m\right)\psi
\nonumber\\
&&+\frac{1}{2}(\partial\sigma)^2-U(\sigma )
-\frac{1}{4}\Omega_{\mu\nu}\Omega^{\mu\nu}
+\frac{1}{2}m^2_\omega\omega^2
-\frac{1}{4}{\vec{\rm R}}_{\mu\nu}{\vec{\rm R}}^{\mu\nu}
+\frac{1}{2}m^2_\rho\vec\rho^{\,2}
-\frac{1}{4}{\rm F}_{\mu\nu}{\rm F}^{\mu\nu}
\nonumber\\
&&-g_\sigma\bar\psi\sigma\psi-
g_\omega\bar\psi\gamma\cdot\omega\psi-
g_\rho\bar\psi\gamma\cdot\vec\rho\vec\tau\psi -
e\bar\psi\gamma\cdot A \frac{(1-\tau_3)}{2}\psi\;.
\label{lagrangian}
\end{eqnarray}
Vectors in isospin space are denoted by arrows, and
bold-faced symbols will indicate vectors in ordinary
three-dimensional space.  The Dirac spinor $\psi$ denotes
the nucleon with mass $m$.  $m_\sigma$, $m_\omega$, and
$m_\rho$ are the masses of the $\sigma$-meson, the
$\omega$-meson, and the $\rho$-meson.  $g_\sigma$,
$g_\omega$, and $g_\rho$ are the corresponding coupling
constants for the mesons to the nucleon. $e^2 /4 \pi =
1/137.036$.  The coupling constants and unknown meson masses
are parameters, adjusted to fit data of nuclear matter and
of finite nuclei. $U(\sigma)$ denotes the non-linear $\sigma$
self-interaction 
\begin{equation}
U(\sigma)~=~\frac{1}{2}m^2_\sigma\sigma^2+\frac{1}{3}g_2\sigma^3+
\frac{1}{4}g_3\sigma^4,
\label{NL}
\end{equation}
and $\Omega^{\mu\nu}$, $\vec R^{\mu\nu}$, and $F^{\mu\nu}$
are field tensors
\begin{eqnarray}
\Omega^{\mu \nu} & = & \partial^{\mu} \omega^{\nu} - 
                       \partial^{\nu} \omega^{\mu} \\
\vec{R}^{\mu \nu} & = & \partial^{\mu} \vec{\rho}^{\,\nu} -
                        \partial^{\nu} \vec{\rho}^{\,\mu}\\
F^{\mu \nu} & = & \partial^{\mu} A^{\nu} - 
                       \partial^{\nu} A^{\mu}.
\end{eqnarray}
The lowest order of the quantum field theory is the {\it
mean-field} approximation: the meson field operators are
replaced by their expectation values.  The $A$ nucleons,
described by a Slater determinant $|\Phi\rangle$ of
single-particle spinors $\psi_i,~(i=1,2,...,A)$, move
independently in the classical meson fields.  The sources
of the meson fields are defined by the nucleon densities
and currents.  The ground state of a nucleus is described
by the stationary self-consistent solution of the coupled
system of Dirac and Klein-Gordon equations.
Due to time reversal invariance, 
there are no currents in the static solution for an even-even
system, and therefore the spatial 
vector components \mbox{\boldmath $\omega,~\rho_3$} and 
${\bf  A}$ of the vector meson fields vanish.
The Dirac equation reads
\begin{equation}
\label{statDirac}
\left\{-i\mbox{\boldmath $\alpha$}
\cdot\mbox{\boldmath $\nabla$}
+\beta(m+g_\sigma \sigma)
+g_\omega \omega^0+g_\rho\tau_3\rho^0_3
+e\frac{(1-\tau_3)}{2} A^0\right\}\psi_i=
\varepsilon_i\psi_i
\end{equation}
The effective mass $m^*({\bf r})$ is defined as
\begin{equation}
\label{effmass}
m^*({\bf r})=m+g_{\sigma}\,\sigma({\bf r}),
\end{equation}
and the potential $V({\bf r})$ as 
\begin{equation}
\label{scapot}
V({\bf r})=g_{\omega}\,\omega^0({\bf r})+
g_{\rho}\,\tau_3\,\rho^0_3({\bf r})
+e{{(1-\tau_3)}\over 2}A^0({\bf r}).
\end{equation}

In nuclei with odd numbers of protons or neutrons time
reversal symmetry is broken. The odd particle induces 
polarization currents and the time-odd components in the
meson fields. These fields play an essential role in the
description of magnetic moments and of moments 
of inertia in rotating nuclei. However,
their effect on deformations and binding energies is
very small and can be neglected to a good 
approximation.

In order to describe ground-state properties of 
open-shell nuclei, pairing correlations have to be taken
into account. For nuclei close to the $\beta$-stability
line, pairing has been included in the relativistic
mean-field model in the form of a simple BCS
approximation \cite{GRT.90}.  However, for nuclei far from
stability the BCS model presents only a poor approximation.
In particular, in drip-line nuclei the Fermi level is found
close to the particle continuum.  The lowest particle-hole
or particle-particle modes are often embedded in the
continuum, and the coupling between bound and continuum
states has to be taken into account explicitly.  The BCS
model does not provide a correct description of the
scattering of nucleonic pairs from bound states to the
positive energy continuum~\cite{DNW.96};
levels high in the continuum become
partially occupied. Including the system in a box of finite
size leads to unreliable predictions for nuclear radii
depending on the size of this box. In the non-relativistic
case, a unified description of mean-field and pairing
correlations is obtained in the 
Hartree-Fock-Bogoliubov (HFB) theory 
\cite{DNW.96}. The ground state of a nucleus $\vert \Phi >$
is represented as the vacuum with respect to
independent quasi-particles. The quasi-particle operators
are defined by a unitary Bogoliubov transformation of the
single-nucleon creation and annihilation operators.  The
generalized single-particle Hamiltonian of HFB theory
contains two average potentials: the self-consistent field
$\hat\Gamma$ which encloses all the long range {\it ph}
correlations, and a pairing field $\hat\Delta$ which sums
up the {\it pp}-correlations. The expectation value of the
nuclear Hamiltonian $< \Phi\vert \hat H \vert \Phi >$ can
be expressed as a function of the hermitian density matrix
$\rho$ and the antisymmetric pairing tensor $\kappa$. The
variation of the energy functional with respect to $\rho$
and $\kappa$ produces the single quasi-particle
Hartree-Fock-Bogoliubov equations
\begin{eqnarray}
\label{equ.2.1}
\left( \matrix{ \hat h - \lambda & \hat\Delta \cr
                -\hat\Delta^* & -\hat h +\lambda
                 } \right) \left( \matrix{ U_k \cr V_k}\right) =
E_k\left( \matrix{ U_k \cr V_k } \right).
\end{eqnarray}
HFB-theory, being a variational approximation, results in a
violation of basic symmetries of the nuclear system. Among
the most important is the non conservation of the
number of particles.  In order that the expectation value
of the particle number operator in the ground state equals
the number of nucleons, equations (\ref{equ.2.1}) contain a
chemical potential $\lambda$ which has to be determined by
the particle number subsidiary condition. The column
vectors denote the quasi-particle wave functions, and $E_k$
are the quasi-particle energies.

If the pairing field $\hat \Delta$ is diagonal and constant, 
HFB reduces to the BCS-approximation. The lower and upper 
components $U_k({\bf r})$ and $V_k({\bf r})$ are proportional, 
with the BCS-occupation amplitudes $u_k$ and $v_k$ as 
proportionality constants. For a more general pairing interaction 
this will no longer be the case. As opposed to the
functions $U_k({\bf r})$, the lower components $V_k({\bf r})$ 
are localized functions of {\bf r}, as long as the
chemical potential $\lambda$ is below the continuum limit.
Since the densities are bilinear products of $V_k({\bf r})$
(see Eqs.(\ref{equ.2.3.e})-(\ref{equ.2.3.h})), the system
is always localized. The HFB wave function can be written
either in the quasiparticle basis as a product of
independent quasi-particle states, or in the {\it canonical basis}
as a highly correlated BCS-state. In the {\it canonical basis}
nucleons occupy  single-particle states. If the chemical potential 
is close to the continuum, the pairing interaction scatters pairs 
of nucleons into continuum states. Because of additional pairing 
correlations, particles which occupy those levels cannot evaporate. 
Mathematically this is expressed by an additional
non-trivial transformation which connects the particle
operators in the canonical basis to the quasi-particles of
the HFB wave function.

The relativistic extension of the HFB theory was introduced
in Ref.~\cite{KR.91}. In the Hartree approximation for
the self-consistent mean field, the Relativistic
Hartree-Bogoliubov (RHB) equations read
\begin{eqnarray}
\label{equ.2.2}
\left( \matrix{ \hat h_D -m- \lambda & \hat\Delta \cr
                -\hat\Delta^* & -\hat h_D + m +\lambda} \right) 
\left( \matrix{ U_k({\bf r}) \cr V_k({\bf r}) } \right) =
E_k\left( \matrix{ U_k({\bf r}) \cr V_k({\bf r}) } \right).
\end{eqnarray}
where $\hat h_D$ is the single-nucleon Dirac
Hamiltonian (\ref{statDirac}), and $m$ is the nucleon mass.
The RHB equations have to be solved self-consistently, with
potentials determined in the mean-field approximation from
solutions of Klein-Gordon equations
\begin{eqnarray}
\label{equ.2.3.a}
\bigl[-\Delta + m_{\sigma}^2\bigr]\,\sigma({\bf r})&=&
-g_{\sigma}\,\rho_s({\bf r})
-g_2\,\sigma^2({\bf r})-g_3\,\sigma^3({\bf r})   \\
\label{equ.2.3.b}
\bigl[-\Delta + m_{\omega}^2\bigr]\,\omega^0({\bf r})&=&
g_{\omega}\,\rho_v({\bf r}) \\
\label{equ.2.3.c}
\bigl[-\Delta + m_{\rho}^2\bigr]\,\rho^0({\bf r})&=&
g_{\rho}\,\rho_3({\bf r}) \\
\label{equ.2.3.d}
-\Delta \, A^0({\bf r})&=&e\,\rho_p({\bf r}).
\end{eqnarray}
for the sigma meson, omega meson, rho meson and photon
field, respectively.  
Due to charge conservation, only the 3rd-component of the 
isovector rho meson contributes. The source terms in 
equations (\ref{equ.2.3.a}) to (\ref{equ.2.3.d}) are 
sums of bilinear products of baryon amplitudes
\begin{eqnarray}
\label{equ.2.3.e}
\rho_s({\bf r})&=&\sum\limits_{E_k > 0} 
V_k^{\dagger}({\bf r})\gamma^0 V_k({\bf r}), \\
\label{equ.2.3.f}
\rho_v({\bf r})&=&\sum\limits_{E_k > 0} 
V_k^{\dagger}({\bf r}) V_k({\bf r}), \\
\label{equ.2.3.g}
\rho_3({\bf r})&=&\sum\limits_{E_k > 0} 
V_k^{\dagger}({\bf r})\tau_3 V_k({\bf r}), \\
\label{equ.2.3.h}
\rho_{\rm em}({\bf r})&=&\sum\limits_{E_k > 0} 
V_k^{\dagger}({\bf r}) {{1-\tau_3}\over 2} V_k({\bf r}),
\end{eqnarray}
where the sums run over all positive energy states. For $M$
degrees of freedom, for example number of nodes on a radial
mesh, the HB equations are $2M$-dimensional and have $2M$
eigenvalues and eigenvectors. To each eigenvector $(U_k,
V_k)$ with eigenvalue $E_k$, there corresponds an
eigenvector $(V^*_k, U^*_k)$ with eigenvalue $-E_k$. Since
baryon quasi-particle operators satisfy fermion commutation
relations, it is forbidden to occupy the levels $E_k$ and
$-E_k$ simultaneously. Usually one chooses the $M$ positive
eigenvalues $E_k$ for the solution that corresponds to a
ground state of a nucleus with even particle number.

The pairing field $\hat\Delta $ in (\ref{equ.2.2}) is an integral
operator with the kernel
\begin{equation}
\label{equ.2.5}
\Delta_{ab} ({\bf r}, {\bf r}') = {1\over 2}\sum\limits_{c,d}
V_{abcd}({\bf r},{\bf r}') {\bf\kappa}_{cd}({\bf r},{\bf r}'),
\end{equation}
where $a,b,c,d$ denote quantum numbers
that specify the Dirac indices of the spinors,
$V_{abcd}({\bf r},{\bf r}')$ are matrix elements of a
general two-body pairing interaction, and the pairing
tensor is defined
\begin{equation}
{\bf\kappa}_{cd}({\bf r},{\bf r}') =
\sum_{E_k>0} U_{ck}^*({\bf r})V_{dk}({\bf r}').
\end{equation}
The integral operator $\hat\Delta$ acts on the wave function
$V_k({\bf r})$:
\begin{equation}
\label{equ.2.4}
(\hat\Delta V_k)({\bf r}) 
= \sum_b \int d^3r' \Delta_{ab} ({\bf r},{\bf r}') V_{bk}({\bf r}'). 
\end{equation}

The self-consistent solution of the
Dirac-Hartree-Bogoliubov integro-differential eigenvalue equations
and Klein-Gordon equations for the meson fields determines the
nuclear ground state. For systems with spherical symmetry, i.e.
single closed-shell nuclei, the coupled system of equations has been
solved using finite element methods in coordinate space
\cite{PVR2.97,PVL.97,LVP.98,LVR.97,VPLR.98}, and by expansion in a basis
of spherical harmonic oscillator~\cite{Gonz.96,LVR.98,VLR.98}. 
For deformed nuclei the present version of the model 
does not include solutions in coordinate space.
The Dirac-Hartree-Bogoliubov equations and the equations for the
meson fields are solved by expanding the nucleon spinors
$U_k({\bf r})$ and $V_k({\bf r})$,
and the meson fields in terms of the eigenfunctions of a
deformed axially symmetric oscillator potential~\cite{GRT.90}.
Of course for nuclei at the drip-lines,
solutions in configurational representation
might not provide an accurate description of properties that
crucially depend on the spatial extension of nucleon densities,
as for example nuclear radii.

The eigensolutions of Eq. (\ref{equ.2.2}) form a set of
orthogonal (normalized) single quasi-particle states. The 
corresponding eigenvalues are the single quasi-particle 
energies. The self-consistent iteration procedure is performed
in the basis of quasi-particle states. The self-consistent 
quasi-particle eigenspectrum is then transformed into the 
canonical basis of single-particle states. The canonical 
basis is defined to be the one in which the matrix
$R_{kk'}=\bigl< V_k({\bf r})\big\vert V_{k'}({\bf r})\bigl>$ is diagonal.
The transformation to the canonical basis determines the energies
and occupation probabilities of single-particle states, which correspond
to the self-consistent solution for the ground state of a nucleus.

The details of ground-state properties of nuclei at the 
drip-lines will depend on the coupling constants and masses of
the effective Lagrangian, on the 
pairing interaction and coupling between bound and
continuum states. Several parameter sets of the mean-field
Lagrangian have been derived that provide a satisfactory description
of nuclear properties along the $\beta$-stability line.
In particular, NL1 \cite{RRM.86}, NL-SH \cite{SNR.93}, and
NL3 \cite{LKR.97}.  The effective forces NL1 and NL-SH
have been frequently used to calculate properties of
nuclear matter and of finite nuclei, and have become
standard parameterizations for relativistic mean-field
calculations.  The parameter set NL3 has been derived more
recently \cite{LKR.97}, by adjusting to ground state properties
of a large number of spherical nuclei.  
Properties calculated with NL3 indicate that this is probably
the best RMF effective interaction so far, both for nuclei
at and away from the line of $\beta$-stability.

In many applications of the relativistic mean-field model
pairing correlations have been taken into account in a very
phenomenological way in the BCS-theory with monopole pairing force
adjusted to the experimental odd-even mass differences. This framework
obviously cannot be applied to the description of the coupling
to the particle continuum in nuclei close to the drip-line.
The question therefore arises, which pairing interaction
$V_{abcd}({\bf r},{\bf r}')$ should be used in Eq. (\ref{equ.2.5}).
In principle, in Ref. \cite{KR.91} a fully relativistic derivation
of the pairing force has been developed, starting from
the Lagrangian (\ref{lagrangian}). Using the Gorkov factorization
technique, it has been possible to demonstrate that
the pairing interaction results from the one-meson exchange
($\sigma$-, $\omega$- and $\rho$-mesons).
In practice, however, it turns out that the
pairing correlations calculated in this way, with coupling
constants taken from standard parameter sets of the RMF model,
are much too strong. The repulsion produced by the
exchange of vector mesons at short distances results in a
pairing gap at the Fermi surface that is by a factor three too large. 
However, as has been argued in many applications of the
Hartree-Fock-Bogoliubov theory, there is no real reason to
use the same effective forces in both the particle-hole and
particle-particle channel. In a first-order approximation,
the effective interaction contained
in the mean-field $\hat\Gamma$ is
a $G$-matrix, the sum over all ladder diagrams. The
effective force in the $pp$ channel, i.e. in the pairing potential
$\hat\Delta$, should be the $K$ matrix, the
sum of all diagrams irreducible in $pp$-direction.
Since very little is known about this matrix in the relativistic
approach, in most applications of the RHB model a
phenomenological pairing interaction has been used, the pairing
part of the Gogny force \cite{BGG.84},
\begin{equation}
V^{pp}(1,2)~=~\sum_{i=1,2}
e^{-(( {\bf r}_1- {\bf r}_2)
/ {\mu_i} )^2}\,
(W_i~+~B_i P^\sigma
-H_i P^\tau -
M_i P^\sigma P^\tau),
\end{equation}
with the set D1S \cite{BGG.84} for the parameters
$\mu_i$, $W_i$, $B_i$, $H_i$ and $M_i$ $(i=1,2)$.
This force has been very carefully adjusted to the pairing
properties of finite nuclei all over the periodic table.
In particular, the basic advantage of the Gogny force
is the finite range, which automatically guarantees a proper
cut-off in momentum space.
The fact that it is a non-relativistic interaction
has little influence on the results of RHB calculations.
A relativistic pairing force should include both a Lorentz-scalar
and a Lorentz-vector part, as for example the interactions
derived in Refs. \cite{KR.91}. However, as opposed
to the mean-field $\Gamma$, where the contribution of the
small components in the Dirac spinors is reflected in a very large
spin-orbit splitting and where the difference between the scalar density
(\ref{equ.2.3.e}) and the vector density (\ref{equ.2.3.f}) leads
to saturation, the contribution of these components in
the pairing channel can be neglected to a very good approximation.
In the pairing channel only levels in the vicinity of the Fermi surface are
really important, or in other words, pairing is a completely
non-relativistic effect.

\begin{figure}[htb] 
\begin{minipage}[t]{80mm}
\rotate[r]
{\epsfig{file=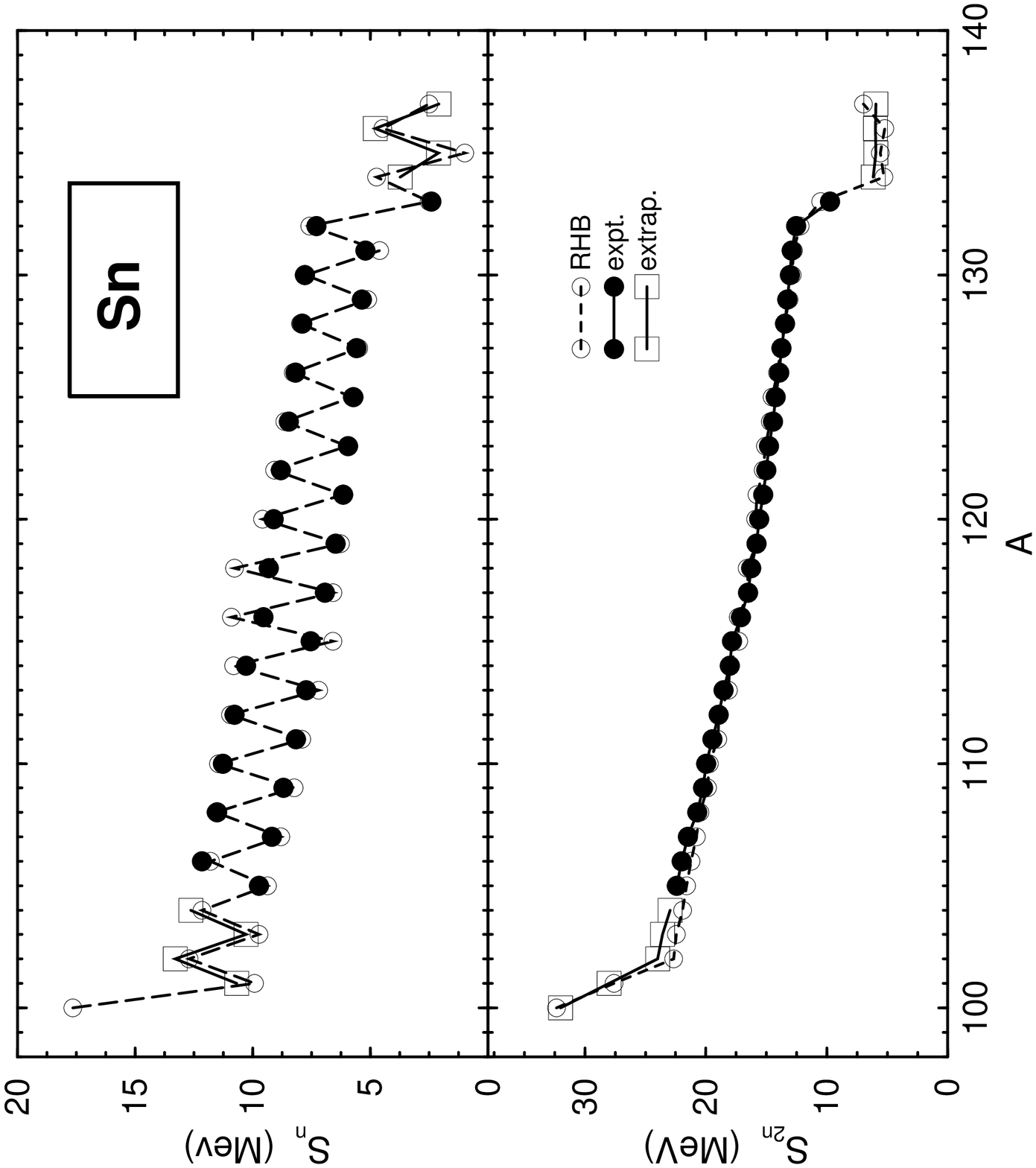,height=75 mm,width=70 mm}}
\vspace{5mm}
\caption{ One and two-neutron separation energies for Sn
isotopes calculated in the RHB model and compared with
experimental (filled circles) and extrapolated (squares)
data from the compilation of G. Audi and A. H. Wapstra.}
\label{fig1}
\end{minipage}
\hspace{0.5 cm}
\begin{minipage}[t]{80mm}
\rotate[r]
{\epsfig{file=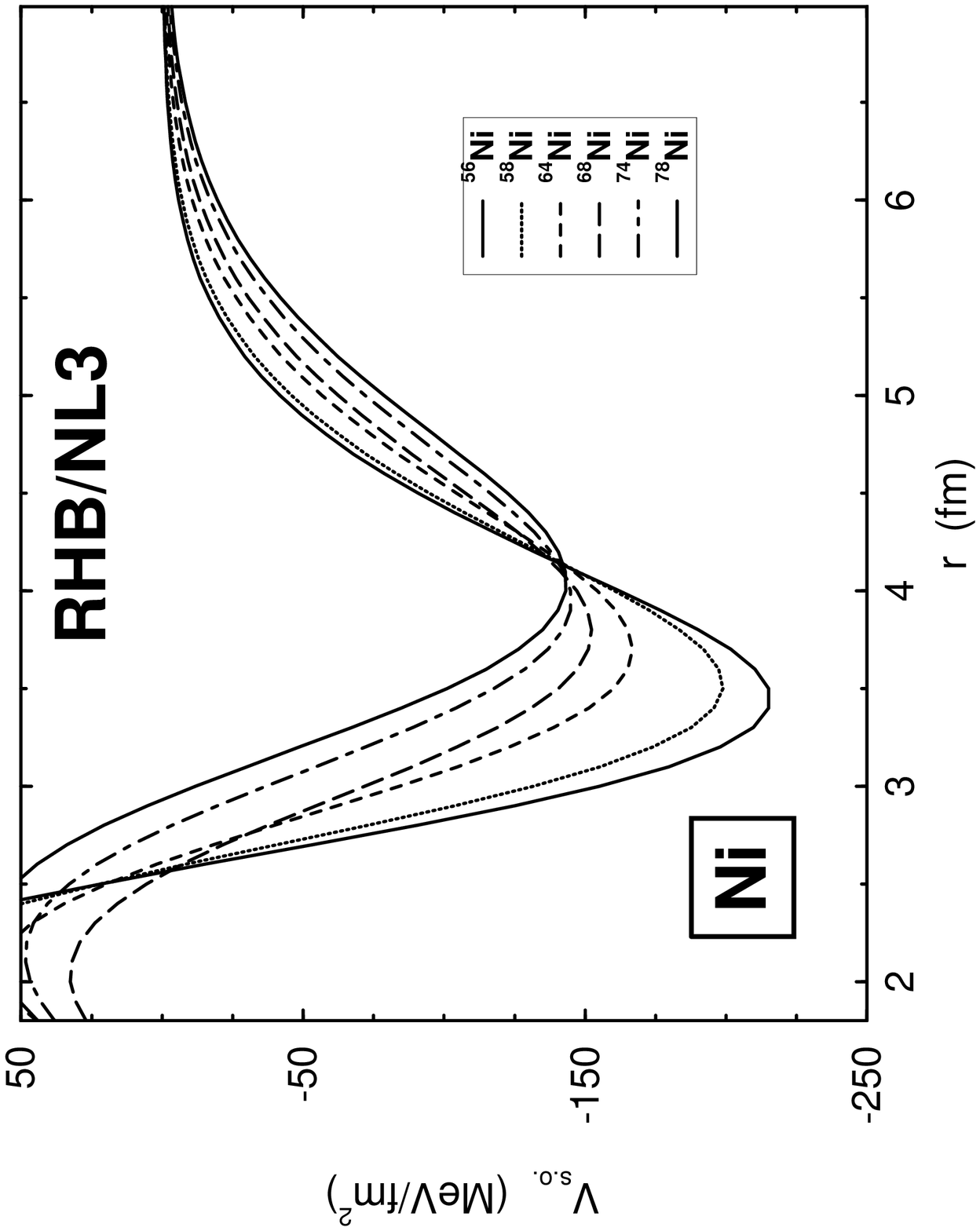,height=79 mm,width=82 mm}}
\vspace{5mm}
\caption{ Radial dependence of the spin-orbit term
of the potential in self-consistent solutions for the
ground-states of Ni ($28 \leq N\leq 50$) nuclei.}
\label{fig2}
\end{minipage}
\end{figure}

In order to test the combination of the NL3 effective interaction for
the mean-field Lagrangian, and the Gogny interaction with the parameter
D1S in the pairing channel, ground-state properties of
a chain of Ni and Sn isotopes have been calculated in Ref. \cite{LVR.98}.
The analysis has shown that the NL3 effective force provides an accurate 
description of systems with very different number of neutrons.
The correct isospin dependence of NL3 implies that
this interaction can be used to make reliable
predictions about drip-line nuclei. 
In Fig. \ref{fig1} the one- and two-neutron separation energies
are shown for the Sn ($50\leq N \leq 88$) isotopes. 
The values that correspond to the self-consistent RHB ground-states are
compared with experimental data and extrapolated values from
Ref. \cite{AW.95}. The theoretical values reproduce in detail the
experimental separation energies.  The model describes not only
the empirical values within one major neutron shell, but it also
reproduces the transitions between major shells. 
The agreement with experimental data is also very good for the 
Ni isotopes. For the total binding energies, 
except for the region around $^{60}$Ni and for $^{100-102}$Sn,
the absolute differences between the calculated and experimental
masses are less than 2 MeV. For Ni the model predicts weaker
binding for $N\leq 40$. Compared to experimental values,
the theoretical binding energies are
$\approx$ 1 MeV larger for neutrons in the 1g$_{9/2}$ orbital
($40\leq N \leq 50$). For Sn isotopes
the results of model calculation in general display
stronger binding. The differences are somewhat larger for $^{100-102}$Sn,
but for these nuclei it is expected that the masses might
be strongly affected by proton-neutron
residual short-range correlations. Excellent results have also been 
obtained in the comparison between calculated and experimental 
neutron and proton rms radii for the Ni and Sn isotopes. 

In Ref. \cite{LR.98} it has been shown that constrained
RMF calculations with the NL3 effective force reproduce the
excitation energies of superdeformed minima relative to the
ground-state in $^{194}$Hg and $^{194}$Pb. In the same work
the NL3 interaction was also used for calculations
of binding energies and deformation parameters of rare-earth
nuclei. All theoretical analyses have shown that the combination
NL3 $+$ D1S could be expected to predict relevant results 
also in the regions of drip-line nuclei. In all the applications 
of the RHB model which are described in the present review, 
the NL3 effective force has been used in the mean-field Lagrangian, 
and the Gogny D1S interaction in the pairing channel.

The spin-orbit interaction plays a central role in 
nuclear structure. It is rooted in the basis of the nuclear
shell model, where its inclusion is essential in order to
reproduce the experimentally established magic numbers.
In non-relativistic models based on the mean-field approximation,
the spin-orbit potential is included in a phenomenological way.
Of course such an ansatz introduces an additional parameter:
the strength of the spin-orbit interaction. The value of this
parameter is usually adjusted to the experimental
spin-orbit splittings in spherical nuclei, for example $^{16}$O.
On the other hand, in the relativistic framework
the nucleons are described as Dirac spinors. This means that in the
relativistic description of the nuclear many-body problem,
the spin-orbit interaction arises naturally from the
Dirac-Lorenz structure of the effective Lagrangian. No
additional strength parameter is necessary, and
relativistic models reproduce the empirical
spin-orbit splittings.  It is then of special interest to study 
the predictions of the relativistic model for the isovector 
properties of the spin-orbit term of the effective interaction \cite{LVR.97}. 
In the first order approximation, and assuming spherical
symmetry, the spin-orbit term can be written as
\begin{equation}
\label{so1}
V_{s.o.} = {1 \over r} {\partial \over \partial r} V_{ls}(r),
\end{equation}
where $V_{ls}$ is the spin-orbit potential~\cite{Rin.96}.
\begin{equation}
\label{so2}
V_{ls} = {m \over m_{eff}} (V-S).
\end{equation}
V and S denote the repulsive vector and the
attractive scalar potentials, respectively.
$m_{eff}$ is the effective mass
\begin{equation}
\label{so3}
m_{eff} = m - {1 \over 2} (V-S).
\end{equation}

Using the vector and scalar potentials from the self-consistent
ground-state solutions, from~(\ref{so1}) - (\ref{so3})  
the corresponding spin-orbit terms have been computed
for the Ni isotopes. They are displayed in Fig. \ref{fig2} as function of
the radial distance from the center of the nucleus.
The magnitude of the spin-orbit term $V_{s.o.}$
decreases with the number of neutrons.
If $^{56}$Ni with $^{78}$Ni are compared,
the reduction is $\approx 35\%$ in the surface region.
This implies a significant weakening of the spin-orbit interaction.
The minimum of $V_{s.o.}$ is also shifted outwards, and this
reflects the larger spatial extension of the scalar and vector densities,
which become very diffuse on the surface. The effect is stronger 
in light nuclei \cite{LVR.97}, while it has been shown 
in Ref. \cite{LVR.98} that the reduction of the
spin-orbit term seems to be less pronounced in the Sn isotopes.
These results indicate that the weakening of the
spin-orbit interaction might not be that important in heavy nuclei.
The effect is reflected in the calculated energy splittings between 
spin-orbit partners. The gradual decrease of the energy gap between
single-particle levels is consistent with the weakening of the
spin-orbit term of the effective interaction.

\section*{Neutron halo in light nuclei}

In some loosely bound systems at the drip-lines,
the neutron density distribution displays an extremely long 
tail: the neutron halo. The resulting large interaction
cross sections have provided the first experimental evidence
for halo nuclei~\cite{Tani.85}. The neutron halo phenomenon
has been studied with a variety of theoretical models~\cite{Tani.95,HJJ.95}.
For very light nuclei in particular, models based on the
separation into core plus valence space nucleons
(three-body Borromean systems) have been employed.
In heavier neutron-rich nuclei one expects that
mean-field models should
provide a better description of ground-state properties.
In a mean-field description, the neutron halo and
the stability against nucleon emission can only be explained
with the inclusion of pairing correlations.
Both the properties of single-particle states near the
neutron Fermi level, and the pairing interaction, are important
in the formation of the neutron halo.

\begin{figure}[htb] 
\begin{minipage}[t]{80mm}
\rotate[r]
{\epsfig{file=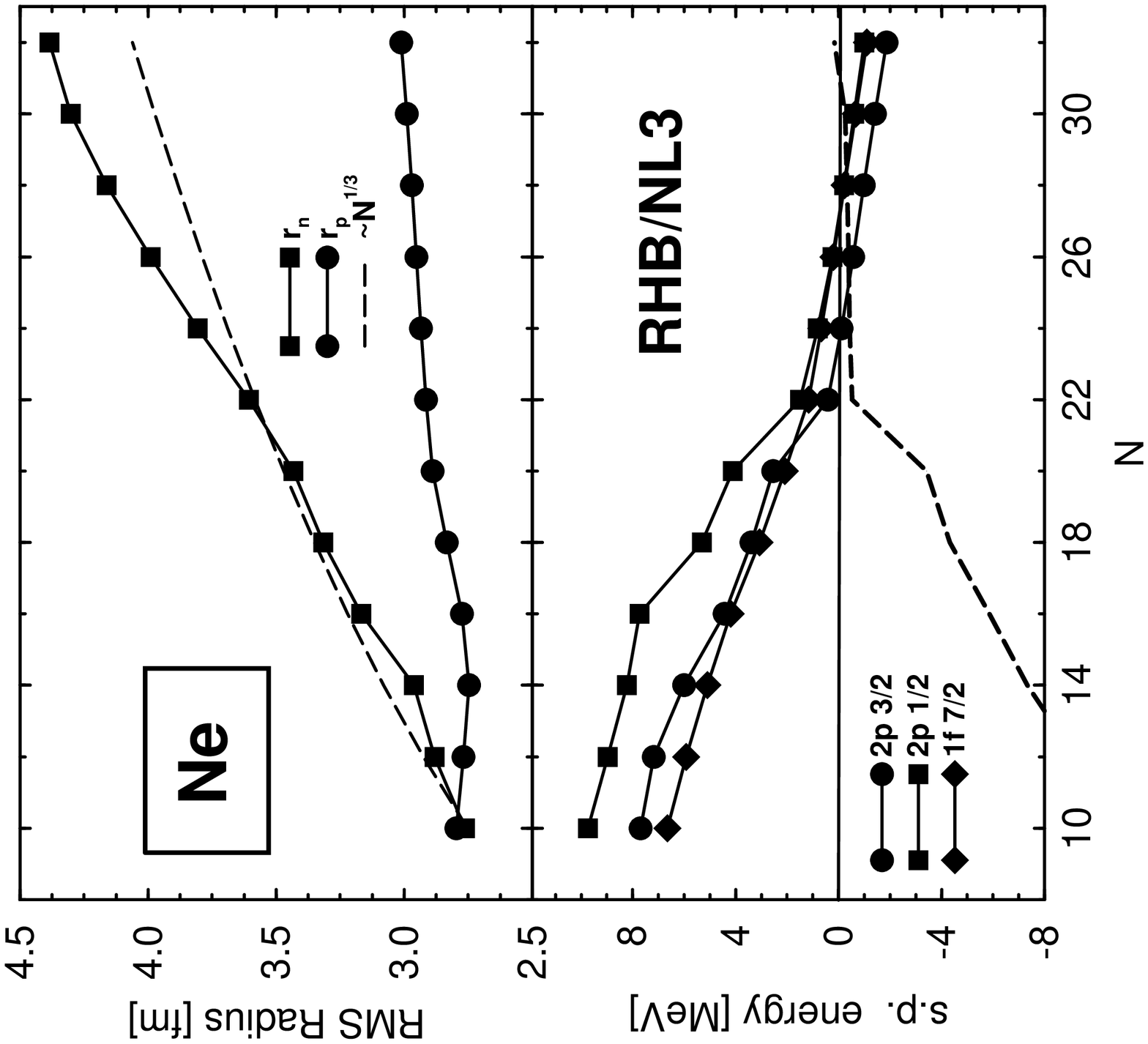,height=80 mm,width=80 mm}}
\vspace{5mm}
\caption{Calculated proton and neutron
$rms$ radii for Ne isotopes (top), and the 1f-2p single-particle neutron
levels in the canonical basis (bottom).}
\label{fig3}
\end{minipage}
\hspace{0.5 cm}
\begin{minipage}[t]{80mm}
\rotate[r]
{\epsfig{file=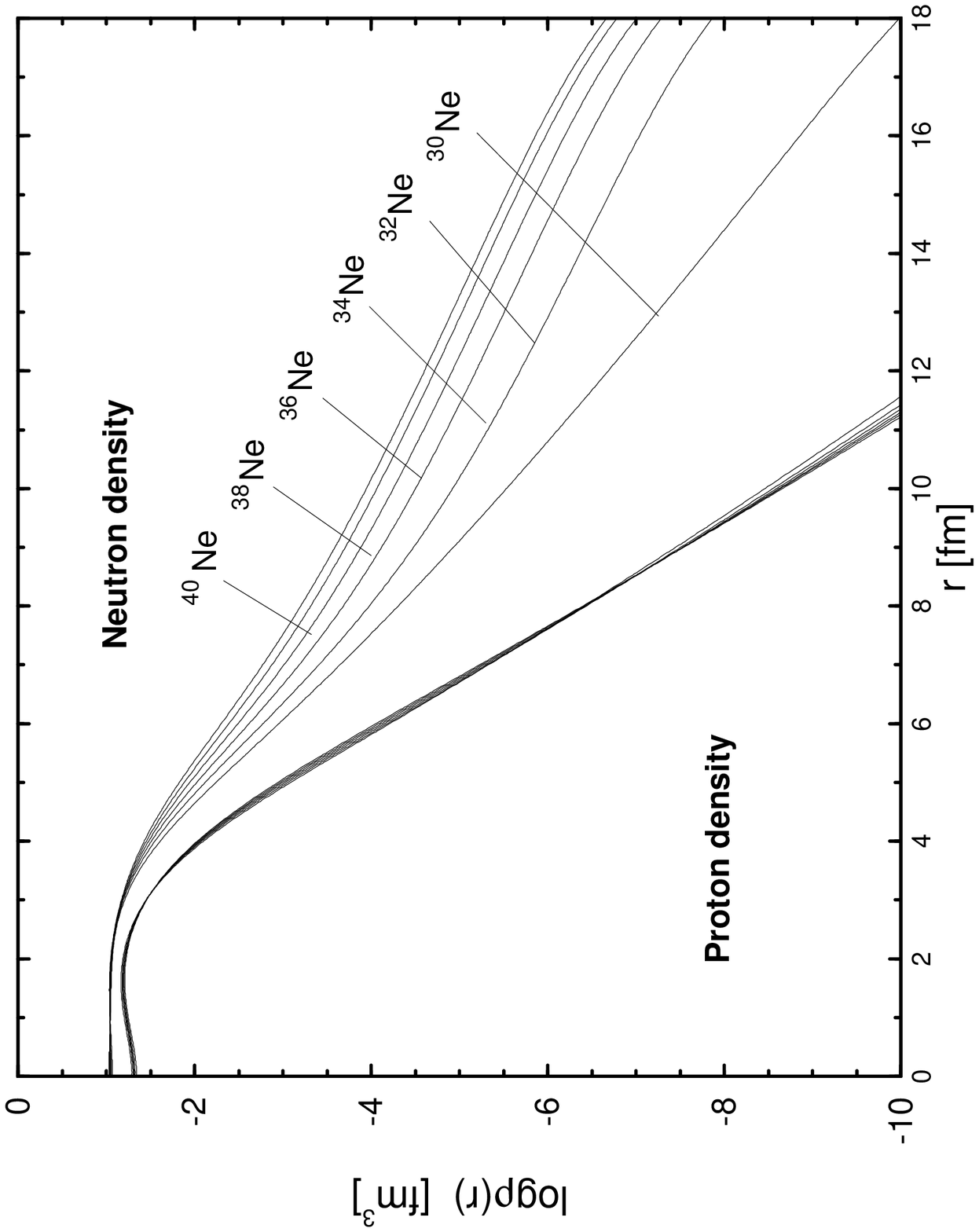,height=80 mm,width=93 mm}}
\vspace{5mm}
\caption{Proton and neutron density distribution for Ne isotopes.} 
\label{fig4}
\end{minipage}
\end{figure}

The details of the formation of the
neutron halo in Ne isotopes have been studied
in Ref.~\cite{PVL.97}. In Fig. \ref{fig3}a the rms radii
for Ne isotopes are plotted as functions of neutron number. 
Neutron and proton rms radii are shown, and the $N^{1/3}$
curve normalized so that it coincides with the neutron
radius in $^{20}$Ne. The neutron radii follow the
mean-field $N^{1/3}$ curve up to  $N \approx 22$.  For
larger values of $N$ the neutron radii display a sharp
increase, while the proton radii stay practically constant.
This sudden increase in neutron rms radii has been
interpreted as evidence for the formation of a
multi-particle halo. The phenomenon is also observed in
the plot of proton and neutron density distributions \ref{fig4}.
The proton density profiles do not change with the number of
neutrons, while the neutron density distributions display
an abrupt change between $^{30}$Ne and $^{32}$Ne.  The
microscopic origin of the neutron halo has been found in a
delicate balance of the self-consistent mean-field and the
pairing field. This is shown in Fig. \ref{fig3}b, where 
the neutron single-particle states 1f$_{7/2}$, 2p$_{3/2}$
and 2p$_{1/2}$ in the canonical basis, and the Fermi
energy are plotted as function of the neutron number.  For $N \leq 22$
the triplet of states is high in the continuum, and the
Fermi level uniformly increases toward zero. The triplet
approaches zero energy, and a gap is formed between these
states and all other states in the continuum. The shell
structure dramatically changes at $N \geq 22$. Between $N =
22$ and $N = 32$ the Fermi level is practically constant and
very close to the continuum. The addition of neutrons in
this region of the drip does not increase the binding. Only
the spatial extension of neutron distribution displays an
increase. The formation of the neutron halo is related to
the quasi-degeneracy of the triplet of states 1f$_{7/2}$,
2p$_{3/2}$ and  2p$_{1/2}$. The pairing interaction
promotes neutrons from the 1f$_{7/2}$ orbital to the 2p
levels. Since these levels are so close in energy, the
total binding energy does not change significantly. Due to
their small centrifugal barrier, the 2p$_{3/2}$ and
2p$_{1/2}$ orbitals form the halo. 
A similar mechanism has been suggested in
Ref.~\cite{MR.96} for the experimentally observed halo in
the nucleus $^{11}$Li. There the formation of the halo
is determined by the pair of neutron levels 1p$_{1/2}$ and
2s$_{1/2}$. A giant halo has been also predicted for
Zirconium isotopes~\cite{MR.98}. In that case the
halo originates from the neutron orbitals
2f$_{7/2}$, 3p$_{3/2}$ and 3p$_{1/2}$.

RHB calculations reported in Ref. \cite{LVP.98}
have shown that the triplet of
single-particle states near the neutron Fermi level:
1f$_{7/2}$, 2p$_{3/2}$ and  2p$_{1/2}$, and the neutron
pairing interaction determine the location of the neutron
drip-line, the formation of the neutron skin, or eventually
of the neutron halo in light nuclei.  For C, N, O and F the triplet 
is still high in the continuum at $N = 20$, and the pairing
interaction is to weak to promote pairs of neutrons into
these levels.  All mean-field effective interactions
predict similar results, and the neutron drip is found at
$N = 18$ or $N = 20$.  For Ne, Na, and Mg the states 1f$_{7/2}$,
2p$_{3/2}$ and  2p$_{1/2}$ are much lower in energy, and
for $N \geq 20$ the neutrons populate these levels. The
neutron drip can change by as much as twelve neutrons. The
model predicts the formation of neutron skin, and
eventually neutron halo in Ne and Na. This is due to the
fact that the triplet of states is almost degenerate in
energy for $N \geq 20$.  For Mg the 1f$_{7/2}$ lies deeper
and neutrons above the $s-d$ shell will exclusively
populate this level, resulting in a deformation of the mean
field.  When the neutron single-particle levels are displayed as
function of the number of protons, it turns out that the
binding of neutrons is determined by the 1f$_{7/2}$ orbital
which is found to be parallel to the slope of the Fermi
level.  The reduction of the spin-orbit splitting
1f$_{7/2}$ - 1f$_{5/2}$ close to the neutron drip-line has
been related to the isospin dependence of the spin-orbit
interaction.  Density distributions have been analyzed, and
the resulting surface thickness and diffuseness parameter
reflect the reduction of the spin-orbit term of the
effective potential in the surface region of neutron-rich
nuclei.

In Ref. \cite{VPLR.98} an interesting study of 
$\Lambda$-hypernuclei with a large neutron excess has 
been reported. In particular, the effects of the
$\Lambda$ hyperon on Ne isotopes
with neutron halo have been studied.
Although the inclusion of the $\Lambda$ does
not produce excessive changes in bulk properties
of these nuclei, it can shift the neutron drip by
stabilizing an otherwise unbound core nucleus at
the drip-line. The microscopic mechanism through which
additional neutrons are bound to the core originates
from the increase in magnitude of the spin-orbit
term in presence of the $\Lambda$ particle. 
The $\Lambda$ in its ground state produces only a
fractional change in the central mean-field potential.
On the other hand, through a purely relativistic
effect, it notably changes the spin-orbit
term in the surface region, providing additional
binding for the outermost neutrons. The effect can be
illustrated on on the example of $^{30}$Ne and the
corresponding hypernucleus $^{31}_{\Lambda}$Ne.
The mean field potential, in which
the nucleons move, results from the cancelation of two
large meson potentials: the attractive scalar potential
S and the repulsive vector potential V: V+S. The spin-orbit
potential, on the other hand, arises from the very strong
anti-nucleon potential V-S. Therefore, while in the presence
of the $\Lambda$ the changes in V and S cancel out in the
mean-field potential, they  are amplified in $V_{ls}$.
For the core $^{30}$Ne
the values of the scalar (S) and vector (V) potential in
the center of the nucleus are -380 MeV and 308 MeV,
respectively. For $^{31}_{\Lambda}$Ne the corresponding
values are: -412 MeV and 336 MeV. The addition of the
$\Lambda$ particle changes the value of the mean-field
potential in the center of the nucleus by 4 MeV, but it
changes the anti-nucleon potential by 60 MeV.
This is reflected in the corresponding spin-orbit term
of the effective potential, which provides more binding for states 
close to the Fermi surface. The additional binding stabilizes
the hypernuclear core.

\section*{Shape coexistence in the deformed $\mathbf{N=28}$ region}

The region of neutron-rich $N\approx 28$ nuclei displays many
interesting phenomena: the average nucleonic potential is
modified, shell effects are suppressed, large quadrupole
deformations are observed as well as shape coexistence,
isovector quadrupole deformations are predicted
at drip-lines. The detailed knowledge of the microscopic structure of
these nuclei is also essential for a correct description of the
nucleosynthesis of the heavy Ca, Ti and Cr isotopes. In the 
the RHB model the structure of exotic neutron rich-nuclei with
$12\leq Z \leq 20$ has been studied, and in particular the light $N=28$
nuclei. Especially interesting is the influence of the spherical shell
$N=28$ on the structure of nuclei below $^{48}$Ca, the deformation
effects that result from the $1f 7/2 \rightarrow fp$ core breaking, and
the shape coexistence phenomena predicted for these $\gamma$-soft nuclei.

The results of Skyrme Hartree-Fock $+$ BCS
and RMF $+$ BCS calculations of Ref. \cite{Wer.96}
have shown that neutron-rich Si, S and Ar isotopes can be considered as
$\gamma$-soft, with deformations depending on subtle interplay between
the deformed gaps $Z=16$ and $18$, and the spherical gap at $N=28$.
Because of cross-shell excitations to the $2p_{3/2}$, $2p_{1/2}$ and
$1f_{5/2}$ shells, the $N=28$ gap appears to be broken in most cases.
In the RMF $+$ BCS analysis of Ref.l\cite{LFS.98}, a careful study of the
phenomenon of shape coexistence was performed for nuclei in this region.
It was shown that several Si and S isotopes exhibit shape coexistence:
two minima with different deformations occur in the binding energy.
The energy difference between the two minima is of the order of
few hundred keV. For $^{44}$S the calculated difference was only
30 keV. The results obtained with the two models, HF and RMF,
were found to be similar, although
also important differences were calculated, as for example, 
the equilibrium shape of the $N=28$ nucleus $^{44}$S. The comparison
with results of FRDM and ETFSI calculations, has also shown the 
importance of a unified and self-consistent description of 
mean-field and pairing correlations in this region of transitional 
nuclei.
\begin{figure}[htb] 
\begin{minipage}[t]{80mm}
\rotate[r]
{\epsfig{file=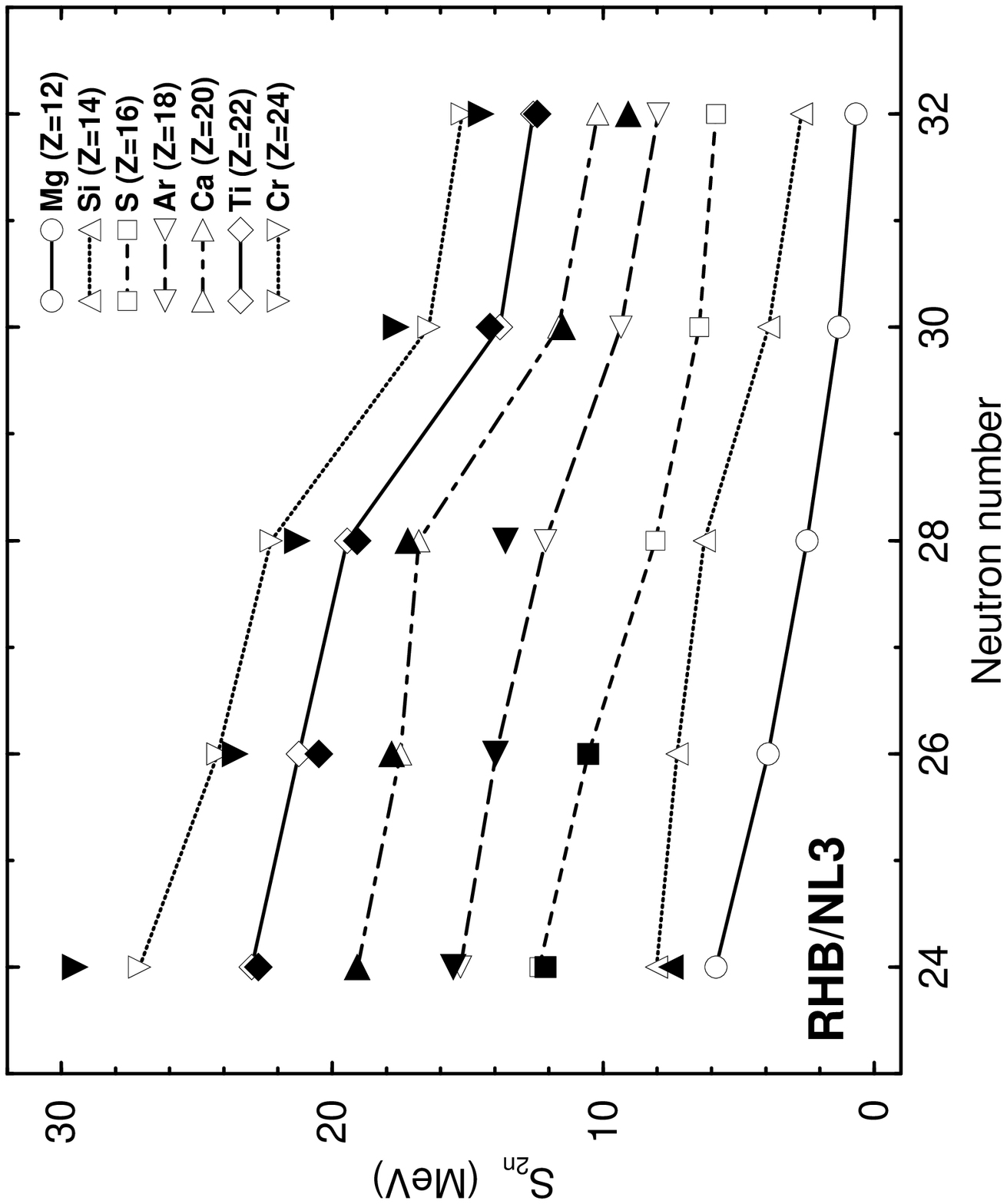,height=80 mm,width=80 mm}}
\vspace{5mm}
\caption{ Two-neutron separation energies in the $N \approx 28$
region calculated in the RHB model and compared with
experimental data (filled symbols)
from the compilation of G. Audi and A. H. Wapstra.}
\label{fig5}
\end{minipage}
\hspace{0.5 cm}
\begin{minipage}[t]{80mm}
\rotate[r]
{\epsfig{file=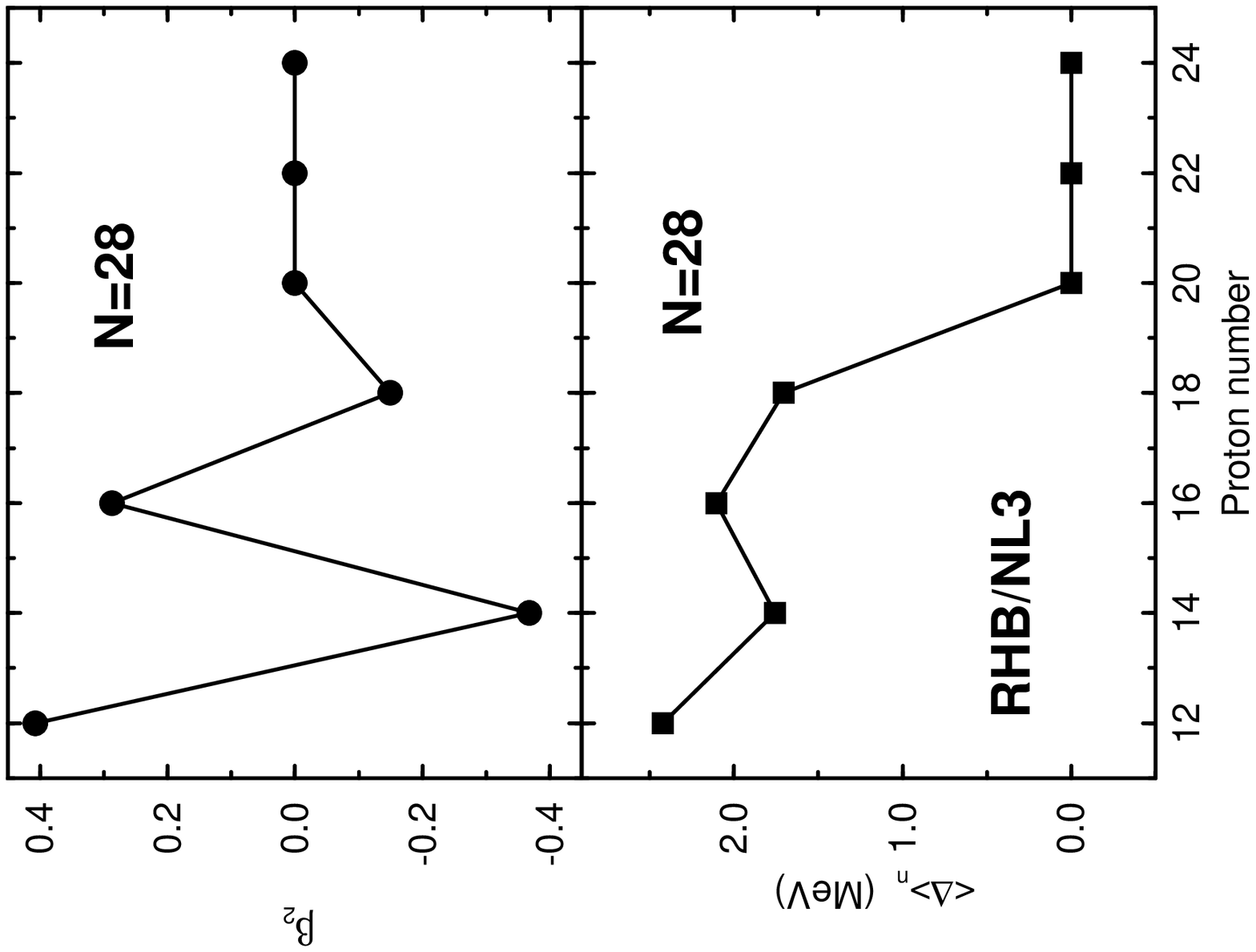,height=80 mm,width=70 mm}}
\vspace{5mm}
\caption{ Self-consistent RHB quadrupole deformations for
ground-states of the $N = 28$ isotones  (top).
Average neutron pairing gaps $< \Delta_N >$
as function of proton number (bottom).}
\label{fig6}
\end{minipage}
\end{figure}

In Fig. \ref{fig5} the two-neutron separation energies are plotted
for the even-even nuclei $12\leq Z \leq 24$ and $24\leq Z \leq 32$.
The values that correspond to the self-consistent RHB ground-states
(symbols connected by lines)
are compared with experimental data and extrapolated values from
Ref.~\cite{AW.95} (filled symbols). Except for Mg and
Si, these isotopes are not at the drip-lines.
The theoretical values
reproduce in detail the experimental separation energies,
except for $^{48}$Cr. In general, it has been found that RHB model
binding energies are in very good agreement with experimental data
when one of the shells (proton or neutron) is closed, or
when valence protons and neutrons occupy different major
shells (i.e. below and above $N$ and/or $Z=20$).
The differences are more pronounced when both protons and neutrons
occupy the same major shell, and especially
for the $N=Z$ nuclei. For these nuclei additional
correlations should be taken into account, and
in particular proton-neutron pairing
could have a strong influence on the masses. 
\begin{figure}[t] 
\rotate[r]
{\epsfig{file=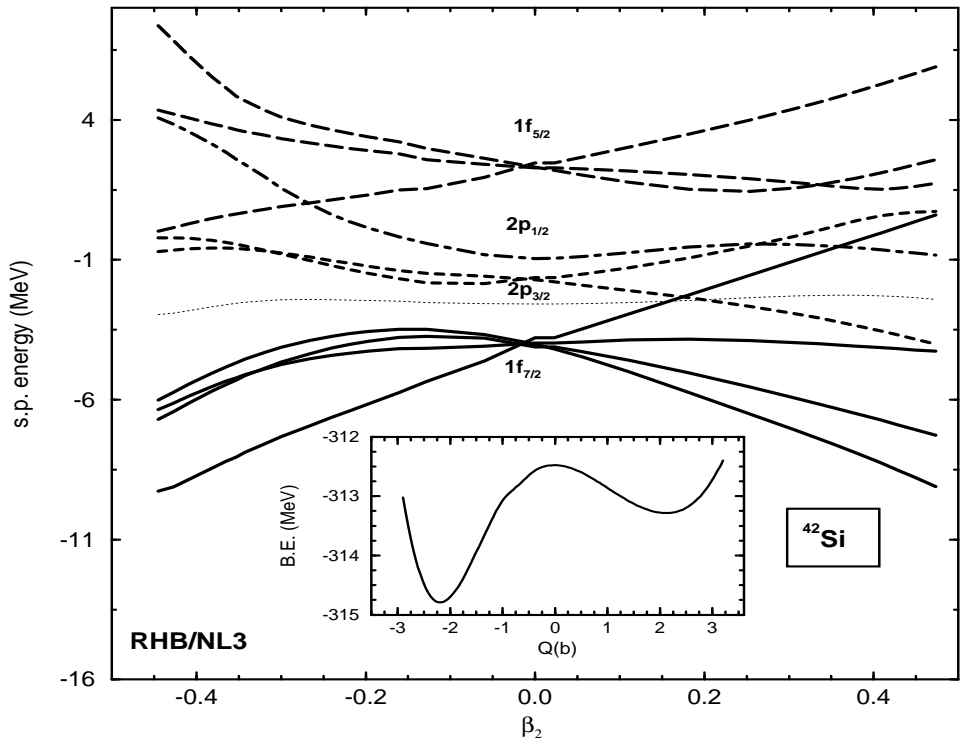,height=140 mm,width=100 mm}}
\vspace{5mm}
\caption{ The neutron single-particle levels for $^{42}$Si as
function of the quadrupole deformation. The energies in the
canonical basis correspond to ground-state RHB solutions
with constrained quadrupole deformation. The dotted line
denotes the neutron Fermi level. In the insert 
the corresponding total binding energy curve is shown.}
\label{fig7}
\end{figure}
\begin{figure}[t] 
\rotate[r]
{\epsfig{file=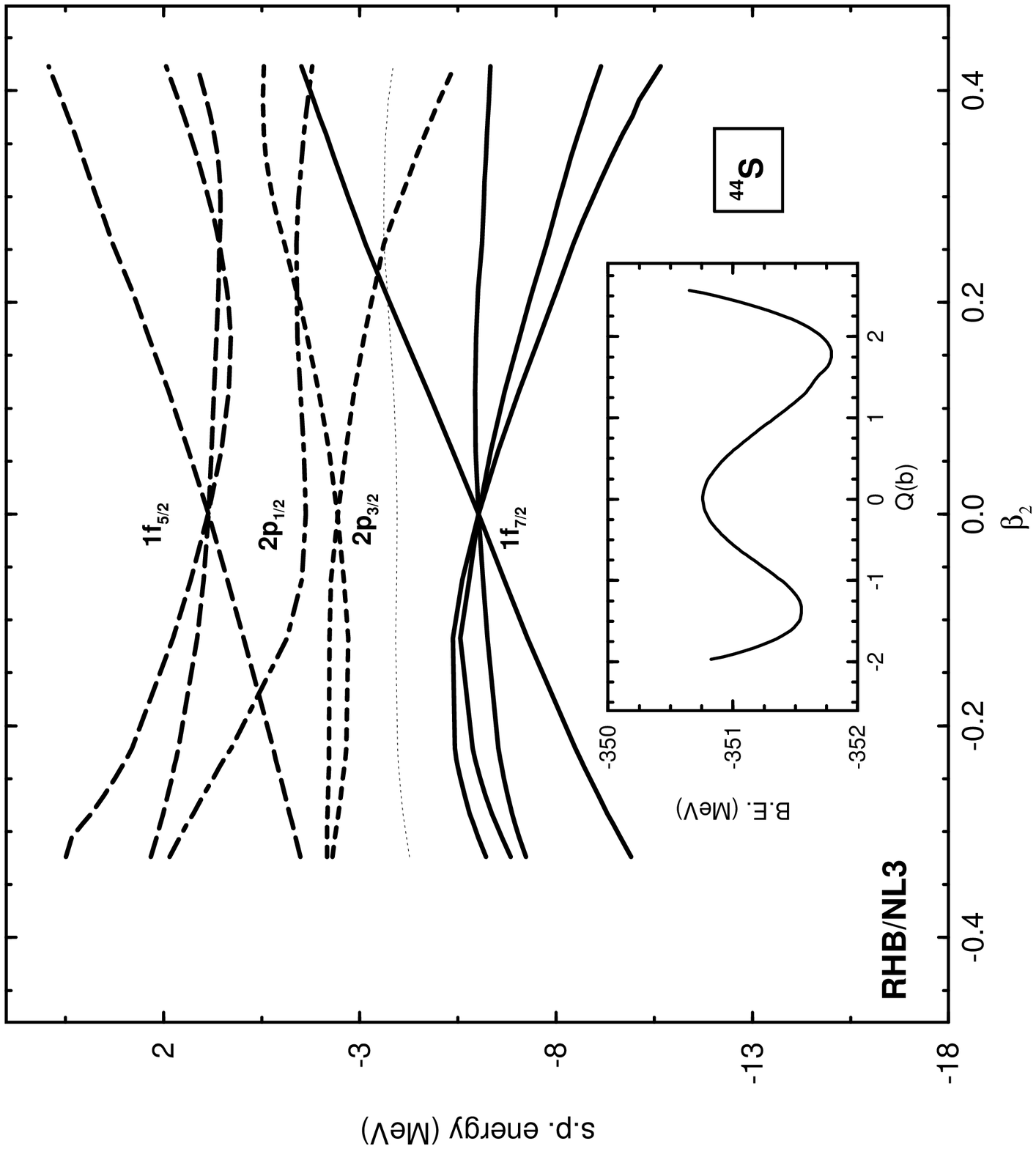,height=140 mm,width=100 mm}}
\vspace{5mm}
\caption{ The same as in Fig. 7, but for $^{44}$S.}
\label{fig8}
\end{figure}

The predicted mass quadrupole deformations for the ground states
of $N=28$ nuclei are shown in the upper panel of Fig. \ref{fig6}.
The staggering between prolate and oblate configurations
indicates that the potential is $\gamma$-soft. The absolute
values of the deformation decrease towards the $Z=20$ closed
shell. Starting with Ca, the $N=28$ nuclei are spherical in the ground
state. The calculated quadrupole deformations are in agreement
with previously reported theoretical results \cite{Wer.96}
(prolate for $Z=16$, oblate for $Z=18$), and with available experimental
data: $| \beta_2 | = 0.258 (36)$ for $^{44}$S ~\cite{Glas.97,Glas.98},
and $| \beta_2 | = 0.176 (17)$ for $^{46}$Ar~\cite{Sch.96}.
Experimental data (energies of $2_1^+$ states and
$B(E2; 0_{g.s.}^+ \rightarrow 2_1^+)$ values) do not determine the
sign of deformation, i.e. do not differentiate between prolate and
oblate shapes. In the lower panel of
Fig. \ref{fig6} the average values of the neutron
pairing gaps for occupied canonical states are displayed.
$< \Delta_N >$ provides an excellent quantitative measure
of pairing correlations. The calculated values of
$< \Delta_N > \approx 2$ MeV correspond to those found in
open-shell Ni and Sn isotopes~\cite{LVR.98}. The spherical
shell closure $N=28$ is strongly suppressed for nuclei
with $Z\leq 18$, and only for $Z\geq 20$ neutron pairing
correlations vanish.

\begin{figure}[htb] 
\rotate[r]
{\epsfig{file=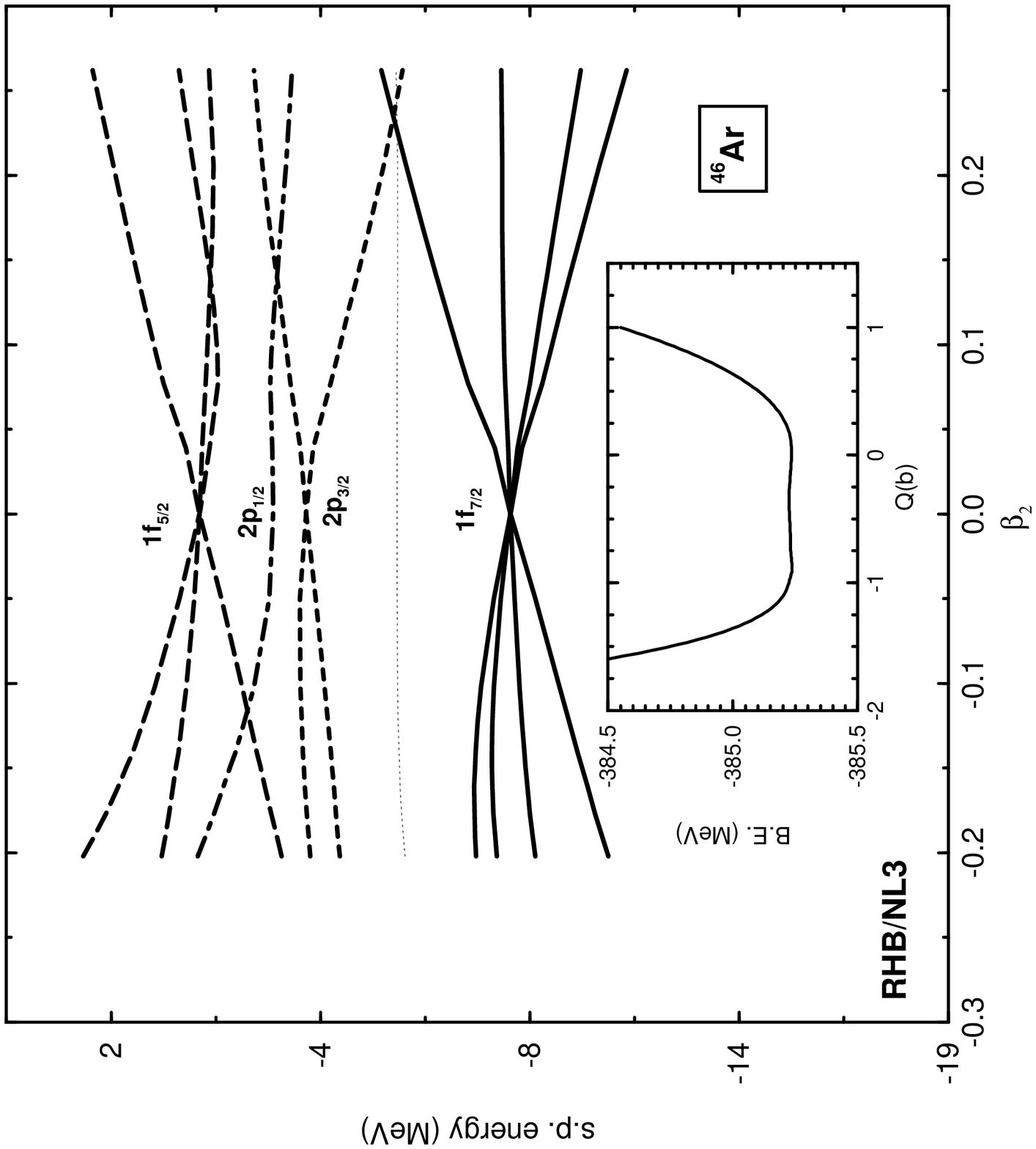,height=140 mm,width=100 mm}}
\vspace{5mm}
\caption{ The same as in Fig. 7, but for $^{46}$Ar.}
\label{fig9}
\end{figure}

In the fully microscopic and self-consistent RHB model, 
the details of the single-neutron levels can be analyzed,
and the formation of minima in the binding energy can be studied. 
Figs.~\ref{fig7}-\ref{fig9} display the single-neutron levels
in the canonical basis for the $N=28$ nuclei $^{42}$Si, $^{44}$S,
and $^{46}$Ar, respectively. The single-neutron eigenstates of the
density matrix result from constrained RHB calculations performed
by imposing a quadratic constraint on the quadrupole moment.
The canonical states are plotted as function of the quadrupole
deformation, and the dotted curve denotes the position of the
Fermi level. In the insert the corresponding total binding
energy curve as function of the quadrupole moment is shown. For $^{42}$Si
the binding energy displays a deep oblate minimum ($\beta_2 \approx -0.4$).
The second, prolate minimum is found at an excitation energy of
$\approx 1.5$ MeV. Shape coexistence is more pronounced for $^{44}$S.
The ground state is prolate deformed, the calculated deformation
in excellent agreement with experimental data~\cite{Glas.97,Glas.98}.
The oblate minimum is found only $\approx 200$ keV above the ground state.
Finally, for the nucleus $^{46}$Ar a very flat energy
surface is found on the oblate side. The deformation of the ground-state
oblate minimum agrees with experimental
data~\cite{Sch.96}, the spherical state is only few keV higher.
It is also interesting to observe how the spherical gap between
the $1f_{7/2}$ orbital and the $2p_{3/2}$, $2p_{1/2}$ orbitals
varies with proton number. While the gap is really strongly reduced
for $^{42}$Si and $^{44}$S, in the $Z=18$ isotone $^{46}$Ar
the spherical gap is $\approx 4$ MeV. Of course from $^{48}$Ca
the $N=28$ nuclei become spherical. Therefore the single-neutron
canonical states in Figs.~\ref{fig7}-\ref{fig9} clearly display
the disappearance of the spherical $N=28$ shell closure for
neutron-rich nuclei below $Z=18$.

\section*{Nuclei at the proton drip-lines}

The decay by direct proton emission
provides the opportunity to study the structure of systems beyond the
drip-line. The phenomenon of ground-state proton radioactivity
is determined by a delicate interplay between the Coulomb and
centrifugal terms of the effective potential. While low-Z
nuclei lying beyond the proton drip-line exist only as short
lived resonances, the relatively high potential energy barrier
enables the observation of ground-state proton emission
from medium-heavy and heavy nuclei. At the drip-lines proton
emission competes with $\beta^+$ decay; for heavy nuclei
also fission or $\alpha$ decay can be favored. The proton
drip-line has been fully mapped up to $Z=21$, and possibly for
odd-Z nuclei up to In \cite{WD.97}. No examples of ground-state
proton emission have been discovered below $Z=50$.

The RHB model with finite range-pairing has been applied in the study
of ground-state properties of spherical even-even nuclei
$14\leq Z \leq 28$ and $N=18,20,22$ \cite{VLR.98}. 
While for these neutron numbers the nuclei with
$14\leq Z \leq 20$ are not really very proton-rich,
nevertheless they are useful for
a comparison of the model calculations with experimental data. 
Of particular interest are the predictions of the model for
the proton-rich nuclei in the 1f$_{7/2}$ region. These nuclei
have recently been extensively investigated in experiments
involving fragmentation of $^{58}$Ni.
The principal motivation of many experimental studies in this region is
the possible occurrence of the two-proton ground-state
radioactivity. In particular, the region around $^{48}$Ni is expected to
contain nuclei which are two-proton emitters. On the other hand,
because of the Coulomb barrier at the proton drip-line,
the emission of a pair of protons may be strongly delayed for
nuclei with small negative two-proton separation energies.

The two-proton separation energies 
that correspond to the self-consistent RHB ground-states
for the even-even nuclei $14\leq Z \leq 28$ and $N=18,20,22$,
have been compared with experimental data and extrapolated values from
Ref. \cite{AW.95}. The theoretical values
reproduce in detail the experimental separation energies,
except for $^{38}$Ca and $^{44}$Ti. 
In Table \ref{TabA} the calculated
total binding energies for the $N=18,20,22$
isotones are displayed in comparison with empirical values. 
As has been already shown in the 
previous section for the neutron rich $N=28$ nuclei, the RHB model
results are in very good agreement with experimental data
when one of the shells (proton or neutron) is closed, or
when valence protons and neutrons occupy different major
shells (i.e. bellow and above $N$ and/or $Z=20$).
The absolute differences between the calculated and experimental
masses are less than 2 MeV.
The differences are larger when both proton and neutron valence
particles (holes) occupy the same major shell, and especially
for the $N=Z$ nuclei $^{36}$Ar and $^{44}$Ti. 
\begin{table*}[htb]
\caption{ Comparison between calculated and empirical
binding energies. All values are in units of MeV; empirical
values are displayed in parentheses.}
\begin{center}
\begin{tabular}{llllll}
$^{32}$Si& 269.02 (271.41)&$^{40}$Ar&343.97 (343.81)&$^{44}$Cr&351.65 (349.99)\\
$^{34}$Si& 284.42 (283.43)&$^{38}$Ca&313.11 (313.04)&$^{46}$Cr&380.19 (381.98)\\
$^{36}$Si& 293.08 (292.02)&$^{40}$Ca&341.99 (342.05)&$^{44}$Fe&312.07 (-)     \\
$^{34}$S & 288.10 (291.84)&$^{42}$Ca&362.95 (361.90)&$^{46}$Fe&352.25 (350.20)\\
$^{36}$S & 307.98 (308.71)&$^{40}$Ti&315.39 (314.49)&$^{48}$Fe&384.42 (385.19)\\
$^{38}$S & 320.77 (321.05)&$^{42}$Ti&348.35 (346.91)&$^{46}$Ni&306.72 (-)     \\
$^{36}$Ar&302.52 (306.71) &$^{44}$Ti&373.15 (375.47)&$^{48}$Ni&349.92 (-)     \\
$^{38}$Ar&327.34 (327.06) &$^{42}$Cr&314.94 (314.20)&$^{50}$Ni&385.52 (385.50)\\
\vspace{2 mm}
\end{tabular}
\end{center}
\label{TabA}
\end{table*}

The RHB results should be also compared with recently reported
self-consistent mean-field calculations of Ref. \cite{Naz.96},
and with properties of proton-rich nuclei calculated with
the shell model \cite{Orm.96}.
The calculations of  Ref. \cite{Naz.96} have been performed
for several mean-field models
(Hartree-Fock, Hartree-Fock-Bogoliubov, and relativistic
mean-field), and for a number of effective interactions.
The results systematically
predict the two-proton drip-line to lie between
$^{42}$Cr and $^{44}$Cr, $^{44}$Fe and $^{46}$Fe,
and $^{48}$Ni and $^{50}$Ni. Recent studies of
proton drip-line nuclei in this region have been
performed in experiments based on $^{58}$Ni
fragmentation on a beryllium target, and evidence has
been reported for particle stability of $^{50}$Ni \cite{Bla.94}.
In the shell-model calculations of Ref. \cite{Orm.96}
absolute binding energies were evaluated by computing
the Coulomb energy shifts between mirror nuclei, and
adding this shift to the experimentally determined binding
energy of the neutron rich isotope. The calculated two-proton
separation energies predicted a proton drip-line in agreement
with experimental data and
with the mean-field results \cite{Naz.96}. Compared to the
RHB results, the shell-model total binding
energies are in somewhat better agreement with experimental data.
However, the two models give almost identical values for
the extracted two-proton separation energies of the
drip-line nuclei. The self-consistent
RHB NL3+D1S two-proton separation energies
at the drip-line are also very close to the values
that result from non-relativistic HFB+Gogny (D1S) calculation
of Ref. \cite{Naz.96}.

Proton radioactivity in odd-Z nuclei has been investigated
in the two spherical regions from $51\leq Z \leq 55$
and $69\leq Z \leq 83$. The systematics of proton decay
spectroscopic factors is consistent with half-lives calculated
in the spherical WKB or DWBA approximations.
Recently reported proton decay rates \cite{Dav.98} indicate
that the missing region of light rare-earth nuclei contains strongly deformed
nuclei at the drip-lines. The lifetimes of deformed proton emitters
provide direct information on the last occupied Nilsson configuration,
and therefore on the shape of the nucleus. Modern models for proton
decay rates from deformed nuclei have only recently been developed.
However, even the most realistic calculations are not
based on a fully microscopic and self-consistent description
of proton unstable nuclei. In particular, such a description
should also include important pairing correlations.

\begin{figure}[htb] 
\begin{minipage}[t]{80mm}
\rotate[r]
{\epsfig{file=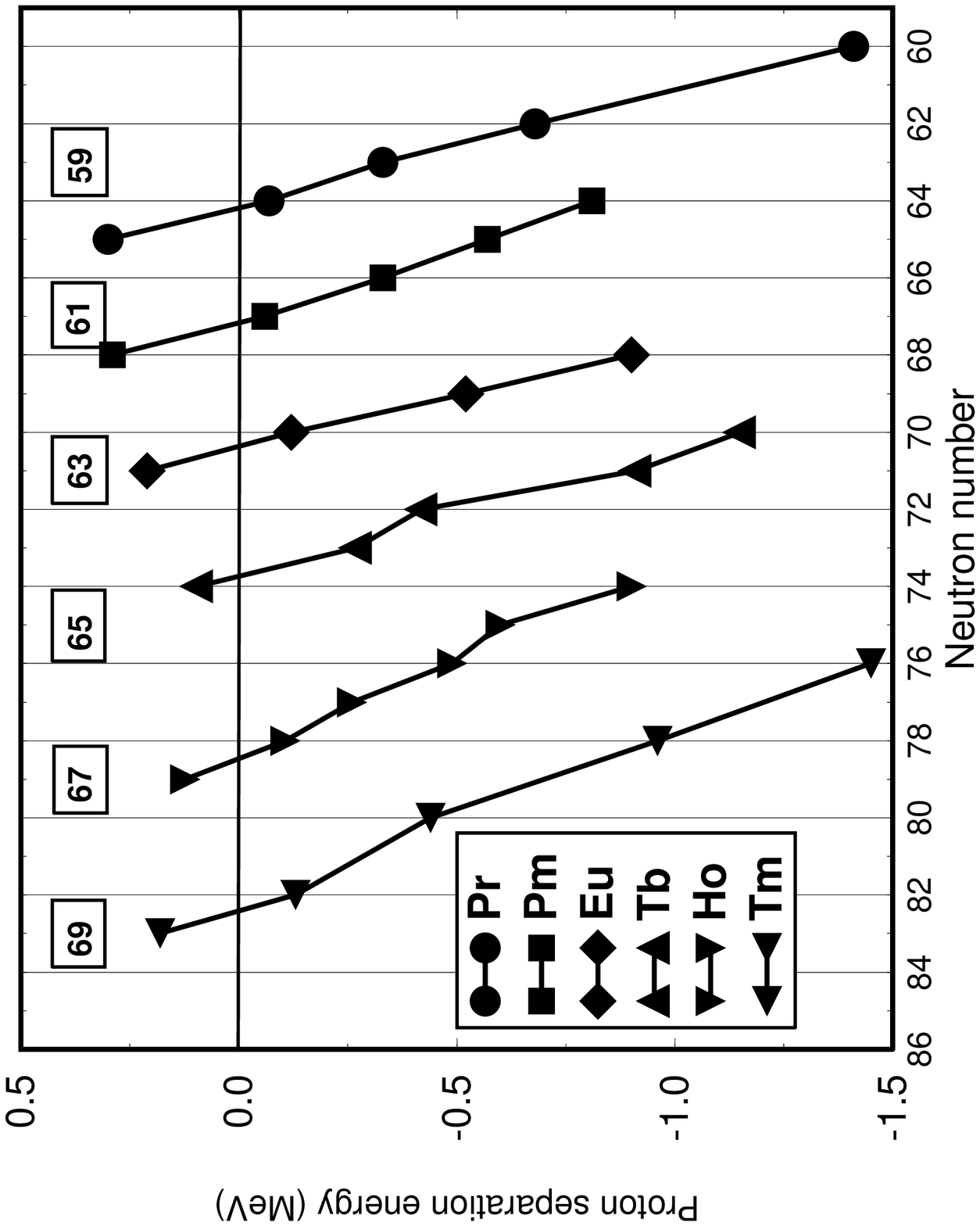,height=80 mm,width=80 mm}}
\vspace{5mm}
\caption{ Calculated one-proton separation energies
for odd-Z nuclei $59 \leq Z \leq 69$ at and beyond the
drip-line.}
\label{fig10}
\end{minipage}
\hspace{0.5 cm}
\begin{minipage}[t]{80mm}
\rotate[r]
{\epsfig{file=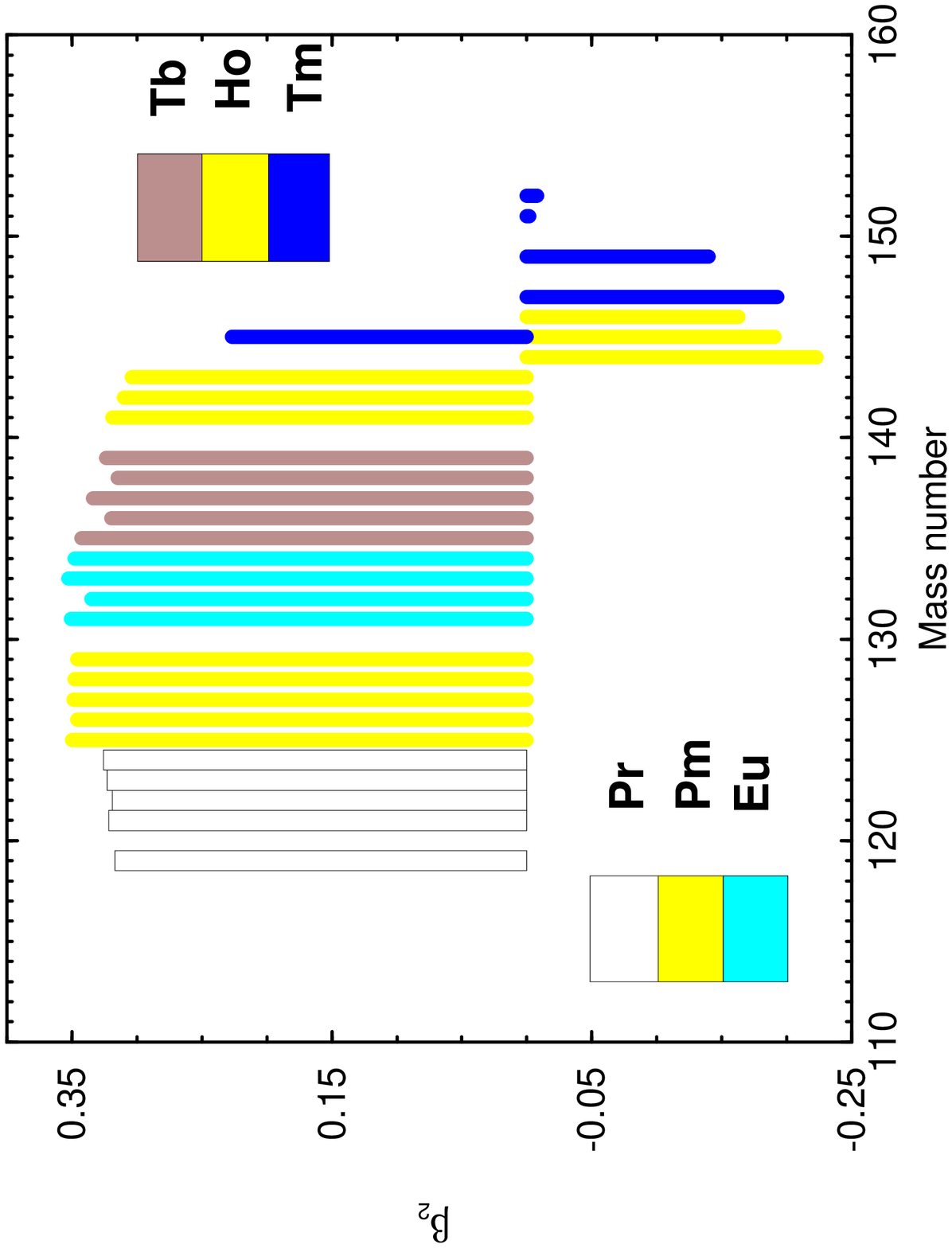,height=80 mm,width=80 mm}}
\vspace{5mm}
\caption{ Self-consistent ground-state quadrupole deformations
for the odd-Z nuclei $59 \leq Z \leq 69$ at the proton drip-line.}
\label{fig11}
\end{minipage}
\end{figure}

Fig. \ref{fig10} displays the one-proton separation energies
for the odd-Z nuclei $59 \leq Z \leq 69$, as function of the
number of neutrons. The model predicts the drip-line nuclei:
$^{124}$Pr, $^{129}$Pm, $^{134}$Eu, $^{139}$Tb, $^{146}$Ho, and
$^{152}$Tm. In heavy proton drip-line nuclei the potential energy
barrier, which results from the superposition of the Coulomb and
centrifugal potentials, is relatively high. For the proton
decay to occur the odd valence proton must penetrate the
potential barrier, and this process competes with $\beta^+$ decay.
Since the half-life for proton decay is inversely proportional
to the energy of the odd proton, in many
nuclei the decay will not be observed immediately after
the drip-line. Proton radioactivity is expected to dominate
over $\beta^+$ decay only when the energy of the odd proton
becomes relatively high. This is also a crucial point for the
relativistic description of proton emitters, since the precise
values of the separation energies depend on
the isovector properties of the spin-orbit interaction.

The calculated separation energies should be compared with
recently reported experimental data on proton radioactivity
from $^{131}$Eu, $^{141}$Ho \cite{Dav.98}, $^{145}$Tm
\cite{Bat.98}, and  $^{147}$Tm \cite{Sel.93}. The $^{131}$Eu
transition has an energy $E_p = 0.950 (8)$ MeV and a half-life
26(6) ms, consistent with decay from either $3/2^+[411]$ or
$5/2^+[413]$ Nilsson orbital. For $^{141}$Ho the transition
energy is $E_p = 1.169 (8)$ MeV, and the half-life 4.2(4) ms
is assigned to the decay of the $7/2^-[523]$ orbital.
The calculated proton separation energy, both for
$^{131}$Eu and $^{141}$Ho, is of $-0.9$ MeV. In
the RHB calculation for $^{131}$Eu the
odd proton occupies the $5/2^+[413]$ orbital, while the
ground state of $^{141}$Ho corresponds to the $7/2^-[523]$
proton orbital.  This orbital is also occupied by the odd
proton in the calculated ground states of $^{145}$Tm and
$^{147}$Tm. The calculated proton separation energies:
$-1.46$ MeV in $^{145}$Tm, and $-0.96$ MeV in $^{147}$Tm, 
are compared with the experimental values for transition
energies: $E_p = 1.728 (10)$ MeV in $^{145}$Tm,
and $E_p = 1.054 (19)$ MeV in $^{147}$Tm. When compared
with spherical WKB or DWBA calculations \cite{ASN.97}, the
experimental half-lives for the two Tm isotopes are consistent
with spectroscopic factors for decays from the $h_{11/2}$ proton
orbital. Though the predicted RHB ground-state configuration
$7/2^-[523]$ indeed originates from the spherical
$h_{11/2}$ orbital, the two nuclei are
deformed. $^{145}$Tm has a prolate quadrupole deformation
$\beta_2 = 0.23$, and $^{147}$Tm is oblate in the ground-state
with $\beta_2 = -0.19$. Calculations also predict possible
proton emitters $^{136}$Tb and $^{135}$Tb with separation
energies $-0.90$ MeV and $-1.15$ MeV, respectively. In both
isotopes the predicted ground-state proton configuration is
$3/2^+[411]$.

The calculated mass quadrupole deformation parameters for
the odd-Z nuclei $59 \leq Z \leq 69$ at and beyond the
drip-line are shown in Fig. \ref{fig11}. Pr, Pm, Eu and
Tb isotopes are strongly prolate deformed
($\beta_2 \approx 0.30 - 0.35$). By increasing the number
of neutrons, Ho and Tm display a transition from prolate to
oblate shapes. The absolute values of $\beta_2$ decrease 
towards the spherical solutions at $N = 82$. The quadrupole
deformations calculated in the RHB model with the NL3
effective interaction, are found in excellent agreement with the
predictions of the macroscopic-microscopic mass model \cite{MN.95}.

\begin{figure}[htb] 
\rotate[r]
{\epsfig{file=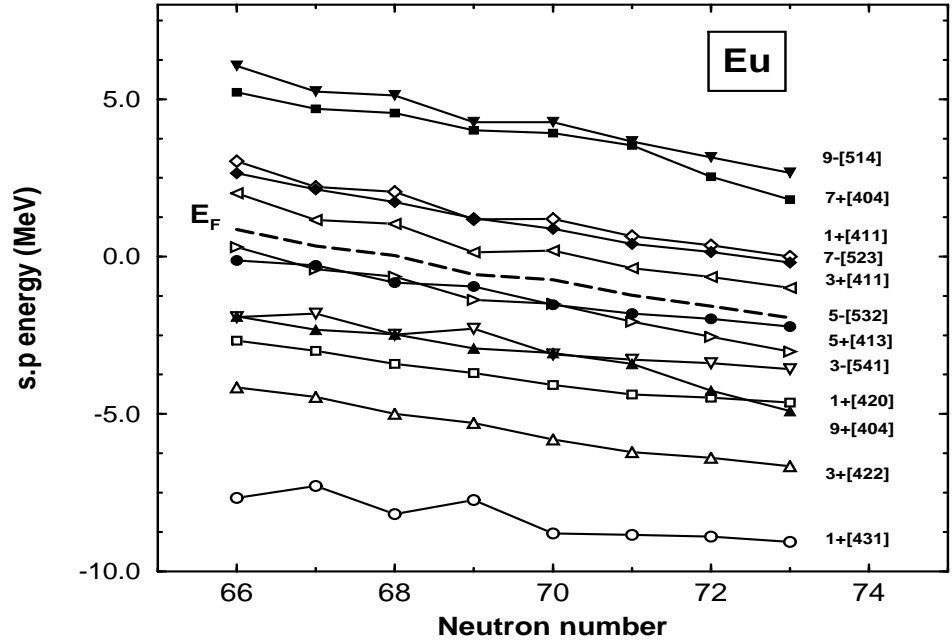,height=140 mm,width=100 mm}}
\vspace{5mm}
\caption{The proton single-particle levels
for the Eu isotopes. The dashed line denotes the position
of the Fermi level. The energies in the canonical basis
correspond to ground-state solutions calculated with the
NL3 effective force of the mean-field
Lagrangian. The parameter set D1S is used for
the finite range Gogny-type interaction in the
pairing channel.}
\label{fig12}
\end{figure}
A detailed analysis of single proton levels, including
spectroscopic factors, can be performed in the canonical basis
which results from the fully microscopic and
self-consistent RHB calculations. For the Eu isotopes
this is illustrated in Fig. \ref{fig12}, where
the proton single-particle energies in the canonical basis
are shown as function of the neutron number.
The thick dashed line denotes the position of the Fermi level.
In particular, for the proton emitter $^{131}$Eu, the ground-state corresponds
to the odd valence proton in the $5/2^+[413]$ orbital.

\section*{Conclusions}
Models based on the relativistic mean-field approximation provide a 
microscopically consistent, and yet simple and economical treatment 
of the nuclear many-body problem. By adjusting just a few model 
parameters, coupling constants and effective masses, to global 
properties of simple, spherical and stable nuclei, it has been 
possible to describe in detail a variety of nuclear structure 
phenomena over the whole periodic table, from light nuclei to 
superheavy elements. When also pairing correlations are included
in the self-consistent Hartree-Bogoliubov framework, the 
relativistic mean-field theory can be applied to the physics 
of exotic nuclei at the drip-lines. In the present review recent
applications of the relativistic Hartree-Bogoliubov model 
have been described: description of the neutron 
drip-line in light nuclei, and of the microscopic mechanism 
for the formation of neutron halos; $\Lambda$-hypernuclei with
large neutron excess; the study of the reduction of the effective
spin-orbit interaction and the resulting 
modification of surface properties in drip-line nuclei; 
the suppression of shell effects and the onset 
of deformation and shape coexistence; the proton drip-line, 
and ground-state proton emitters in the region of deformed 
rare-earth nuclei.

The excellent results reported in these studies clearly reflect 
the basic advantages of the relativistic models over the more
traditional non-relativistic approaches to nuclear structure:
(i) models based on quantum hadrodynamics are more fundamental 
and they explicitly include mesonic degrees of freedom; (ii)
they incorporate important relativistic effects, for instance
the strong scalar and vector potentials and the resulting spin-orbit 
interaction. Of course, at the present stage, the most successful
relativistic models are phenomenological. Although important 
results have been reported in the microscopic derivation 
of the low-energy hadronic Lagrangian,
it is still not possible to bridge the gap between QCD as the 
underlying theory of the strong interaction, and the effective 
theories that have to be used in the nuclear many-body problem. 
In applications to nuclear structure phenomena, one would also
like to see a more consistent treatment of pairing correlations.
The relativistic theory of pairing presents a very active area 
of research. However, only phenomenological effective forces
have been shown to produce reliable results when 
applied to finite nuclei, especially in exotic regions. Important 
improvements should be expected also from models that include density 
dependent interactions and describe nucleons as composite objects. 
Particularly interesting are the extension of the RPA to the relativistic
framework \cite{MGT.97}, and the time-dependent relativistic mean-field
model \cite{VBR.95,Vre.97}, which can be used to describe excited
collective states and analyze recent data on giant resonances at drip-lines.
Large amplitude collective motion in exotic nuclei should also 
provide an excellent example for studies of 
the transition from regular to chaotic dynamics in quantum 
systems \cite{Vre.97a}.

\end{document}